\documentclass[fleqn,usenatbib]{mnras}

\usepackage{newtxtext,newtxmath}

\usepackage[T1]{fontenc}

\usepackage{graphicx}
\usepackage[skip=0pt]{caption}
\usepackage{amsmath}	
\usepackage{colortbl}
\usepackage{xspace}

\newcommand{\result}[1]{\textcolor{black}{#1}}

\newcommand{\cmc}{\textsc{cmc}\xspace}

\newcommand{\msun}{\ensuremath{M_\odot}\xspace}

\title[Gravitational-wave inference of globular clusters]{Globular cluster formation histories, masses and radii inferred from gravitational waves}

\author[Fishbach \& Fragione]{
Maya Fishbach$^{1, 2, 3, 4}$
and Giacomo Fragione$^{4, 5}$
\\
$^{1}$Canadian Institute for Theoretical Astrophysics, 60 St George St, University of Toronto, Toronto, ON M5S 3H8, Canada\\
$^{2}$David A. Dunlap Department of
Astronomy and Astrophysics, 50 St George St, University of Toronto, Toronto, ON M5S 3H8, Canada\\
$^{3}$Department of Physics, 60 St George St, University of Toronto, Toronto, ON M5S 3H8, Canada\\
$^{4}$Center for Interdisciplinary Exploration and Research in Astrophysics (CIERA), 1800 Sherman, Evanston, IL 60201, USA\\
$^{5}$Department of Physics \& Astronomy, Northwestern University, Evanston, IL 60208, USA
}

\date{Accepted XXX. Received YYY; in original form ZZZ}

\pubyear{2023}

\begin{document}
\label{firstpage}
\pagerange{\pageref{firstpage}--\pageref{lastpage}}
\maketitle

\begin{abstract}
    Globular clusters (GCs) are found in all types of galaxies and harbor some of the most extreme stellar systems, including black holes that may dynamically assemble into merging binaries (BBHs). 
    Uncertain GC properties, including when they formed, their initial masses and sizes, affect their production rate of BBH mergers. 
    Using the gravitational-wave catalog GWTC-3, we measure that dynamically-assembled BBHs -- those that are consistent with isotropic spin directions -- make up $\result{61^{+29}_{-44}\%}$ of the total merger rate, with a local merger rate of $\result{10.9^{+16.8}_{-9.3}}$ Gpc$^{-3}$ yr$^{-1}$ rising to $\result{58.9^{+149.4}_{-46.0}}$ Gpc$^{-3}$ yr$^{-1}$ at $z = 1$. 
    We assume this inferred rate describes the contribution from GCs and compare it against the Cluster Monte Carlo (\cmc) simulation catalog to directly fit for the GC initial mass function, virial radius distribution, and formation history.
    We find that GC initial masses are consistent with a Schechter function with slope $\result{\beta_m = -1.9^{+0.8}_{-0.8}}$. Assuming a mass function slope of $\beta_m = -2$ and a mass range between $10^4$--$10^8\,\msun$, we infer a GC formation rate at $z = 2$ of $\result{5.0^{+9.4}_{-4.0}}$ Gpc$^{-3}$ yr$^{-1}$, or $\result{2.1^{+3.9}_{-1.7}}\times 10^6\,\msun$ Gpc$^{-3}$ yr$^{-1}$ in terms of mass density. We find that the GC formation rate probably rises more steeply than the global star formation rate between $z = 0$ and $z = 3$ (\result{82\%} credibility) and implies a local number density that is $\result{f_\mathrm{ev} = 22.6^{+29.9}_{-16.2}}$ times higher than the observed density of survived GCs. This is consistent with expectations for cluster evaporation, but may suggest that other environments contribute to the rate of BBH mergers with significantly tilted spins.
\end{abstract}




\section{Introduction}
Globular clusters (GCs) -- gravitationally-bound, compact collections of stars -- play a key role in astrophysics, from their relationship to galaxy assembly to the exotic stellar populations they generate~\citep[see reviews by][]{1979ARA&A..17..241H,2006ARA&A..44..193B,2018RSPSA.47470616F,2019A&ARv..27....8G}. These exotic systems, which arise from dynamical interactions in GCs, include X-ray binaries, pulsars, fast radio bursts, and potentially, merging binary black hole (BBH) systems~\citep[e.g.,][]{1993Natur.364..423S,IvanovaHeinke2008,GieslerClausen2018,YeKremer2019,KirstenMarcote2022}. Studying these populations can constrain the highly uncertain formation and evolution of GCs themselves, complimentary to direct observations of GCs and high-redshift proto-clusters~\citep{2002ApJ...573...60C,2017MNRAS.469L..63R,2017MNRAS.467.4304V,2022ApJ...940L..53V,2022ApJ...937L..35M}. 

The gravitational-wave (GW) detector network consisting of Advanced LIGO~\citep{2015CQGra..32g4001L}, Advanced Virgo~\citep{2015CQGra..32b4001A} and KAGRA~\citep{2021PTEP.2021eA101A} has observed dozens of BBH mergers to date, and the LIGO-Virgo-KAGRA (LVK) Collaboration reported their latest observations in the Third Gravitational-Wave Transient Catalog (GWTC-3). These observations include 69 confident BBH\footnote{We classify systems in which both component masses are greater than $3\,M_\odot$ as BBH.} events with a false alarm rate FAR $< 1$ yr$^{-1}$~\citep{2021arXiv211103606T,2021SoftX..1300658A}. Although we cannot confidently determine whether any given BBH system originated from a GC, the full BBH population is a promising probe of GC properties out to high redshifts. At their current sensitivities, GW detectors can observe BBH mergers out to redshifts $z \gtrsim 1$, with planned detector upgrades extending this to $z \gtrsim 2$ in the next few years, and proposed next-generation detectors reaching $z \gtrsim 15$~\citep{2018LRR....21....3A,2020JCAP...03..050M,2021arXiv210909882E}. Because we expect BBH mergers to experience a delay time of up to several Gyr between formation and merger, GW detections at redshifts $z < 1$ are still sensitive to GC properties past $z > 2$~\citep{2022arXiv220610622C}. 

The rate of BBH mergers originating from GCs carries information about the GC abundance as well as their mass and radius distribution, because more massive and compact GCs give rise to more BBH mergers. 
Moreover, the BBH merger rate as a function of redshift depends on when GCs formed.
Different groups have explored how varying assumptions about the GC formation histories, masses and radii affects the predicted BBH merger rate~\citep{AskarSzkudlarek2017,2018ApJ...866L...5R,2020PhRvD.102l3016A,2020ApJS..247...48K}.
Recently, \citet{2021MNRAS.506.2362R} showed that next-generation GW detectors can constrain GC formation redshifts by measuring the rate of BBH mergers that originate from GCs out to $z > 15$. This measurement in turn probes the connection between GCs and cosmic reionization. If the bulk of GC star formation occurred at very high redshifts $z \gtrsim 6$, they likely played a significant role in reionization~\citep{1968ApJ...154..891P}. On the other hand, if star formation in GCs largely traces the global star formation rate (SFR), we expect it to peak closer to the peak of the SFR at $z\sim2$~\citep{2019MNRAS.482.4528E,2021MNRAS.503...31T}. 

In this work, we show that GW observations can already provide interesting constraints on GC properties, including their formation histories, masses and virial radii. As described in \S\ref{sec:GCpred}, we use the \cmc models from~\citet{2020ApJS..247...48K} to predict the BBH merger rate as a function of the GC population. \citet{2021RNAAS...5...19R} recently argued that default assumptions about the GC population predict a BBH merger rate consistent with their total rate between $z = 0$ and $z = 1$ as measured from GWTC-2 ~\citep{2021PhRvX..11b1053A,2021ApJ...913L...7A}. Nevertheless, GCs are unlikely to account for the total measured BBH merger rate~\citep{2021ApJ...910..152Z,2021PhRvD.103h3021W,2022MNRAS.511.5797M}. The clearest indication for this is that BBH mergers that are dynamically assembled in GCs are expected to have isotropically-oriented spin directions~\citep{2016ApJ...832L...2R}, while only a fraction of BBH systems are consistent with isotropic spins~\citep{2021ApJ...913L...7A,2021arXiv211103634T,2022ApJ...937L..13C,2022PhRvD.106j3019T,2022ApJ...935L..26F}. Therefore, we first use GWTC-3 to measure the rate of BBH mergers that are consistent with a GC origin in \S\ref{sec:GW-fit}. We do this by fitting the BBH spin distribution in addition to the merger rate as a function of redshift. In \S\ref{sec:GC-fit}, we then use this dynamically-assembled BBH merger rate and its evolution between $z = 0$ and $z = 1$ to infer GC population properties. We fit the GC formation rate as a function of redshift, including the contribution from cluster evaporation, the GC virial radius distribution, and the GC mass function. We discuss directions for future work in \S\ref{sec:discussion} and conclude by summarizing our results in \S\ref{sec:conclusion}. The Appendix contains additional analysis details and corner plots showing hyper-posteriors. Throughout this work, all reported credible regions correspond to median values with 90\% symmetric credible bounds. Our analysis code and data behind the figures can be found at \url{https://github.com/mfishbach/dynamical-bbh-evolution}.

\section{Globular cluster predictions}
\label{sec:GCpred}

This section describes the GC simulations (\S\ref{sec:models}) and population distributions (\S\ref{sec:GC-param}).

\subsection{Globular cluster models}
\label{sec:models}

To estimate the population of merging BBHs in GCs, we use the cluster catalog models of \citet{2020ApJS..247...48K}, run with the \cmc Monte Carlo $N$-body code \citep[][]{RodriguezWeatherford2022}. The models we use span a wide range of initial conditions, including different initial numbers of stars ($N = 2\times10^5$, $4\times10^5$, $8\times10^5$, $1.6\times10^6$), corresponding to stellar masses ($M/\msun =1.2\times10^5$, $2.4\times10^5$, $4.8\times10^5$, $9.6\times10^5$), virial radii ($r_v/\rm{pc} = 0.5, 1, 2, 4$), metallicities ($Z = 0.0002, 0.002, 0.02$), and Galactocentric distances ($R_g/\rm{kpc}=2, 8, 20$)\footnote{The \cmc catalog covers this grid with the exception of the three runs corresponding to $N = 1.6\times10^6$, $r_{\rm v}/\rm{pc} = 0.5$ and $Z = 0.0002$, which led to a runaway process. In order to have a uniform grid of GC parameters, we artificially replace these failed runs with the higher-metallicity, $N = 1.6\times10^6$, $r_v/\rm{pc} = 0.5$, $Z = 0.002$ models. Based on the small difference in the BBH mass spectrum and merger rates between other $Z = 0.0002$ versus $Z = 0.002$ models, we estimate this introduces a $< 10\%$ error to our results.}.

{The star cluster models are assumed to be initially described by a King profile \citep{King1966} with a concentration parameter $W_0=5$, with individual stellar masses sampled from a Kroupa initial mass function between $0.08$ and $150~\msun$ \citep{Kroupa2001}. All models are assumed to have a $5\%$ primordial binary fraction, with secondary masses drawn from a uniform distribution in mass
ratio \citep[e.g.,][]{DuquennoyMayor1991}; binary orbital periods are sampled from a log-uniform distribution from near contact to the hard/soft boundary, while eccentricities are drawn from a thermal distribution \citep{Heggie1975}.}

{Single and binary stars are evolved with the \textsc{SSE} and \textsc{BSE} codes, respectively~\citep{HurleyPols2000,HurleyTout2002}, with up-to-date prescriptions for neutron star and BH formation \citep{FryerBelczynski2012,BelczynskiHeger2016}. 
Natal kicks are assumed to be drawn from a Maxwellian distribution with a standard deviation of $265\,\rm{km \,s^{-1}}$ for NSs formed in core-collapse supernovae~\citep{HobbsLorimer2005}, while BH kicks are reduced in magnitude according to the fractional mass of the supernova fallback material~\citep{FryerBelczynski2012}.
}

{Each simulation is evolved up to a final time of $14$\,Gyr, unless the cluster disrupts or undergoes a collisional runaway process. The \textsc{CMC} cluster catalog model reproduces well the observed properties of the population of Galactic GCs, including their masses, densities, and ages \citep{2020ApJS..247...48K}.}

\subsection{Globular cluster population}
\label{sec:GC-param}
We describe the GC population as a differential rate density over cluster masses $M$, virial radii $r_v$, and metallicities $Z$ as a function of formation redshift $z_\mathrm{form}$. We assume that the formation mass, radius and redshift are all independently distributed, and the metallicity distribution depends on redshift:
\begin{align}
    \frac{d{N}_\mathrm{GC}}{dV_c dt dM dr_v dZ}&\bigl(z_\mathrm{form}\bigr)  = \nonumber \\ &p(M)p(r_v)p(Z\mid z_\mathrm{form})\mathcal{R}_\mathrm{GC}(z_\mathrm{form}),
\end{align}
where $V_c$ is comoving volume and $t$ is time (as measured in the source frame).
At a given merger redshift $z_m$, each \cmc simulation at mass--virial radius--metallicity gridpoint $(M^i, {r_v}^j, Z^k)$ with grid spacing $({\Delta_M}^i, {\Delta_r}^j, {\Delta_Z}^k)$ produces the rate of BBH mergers:
\begin{align}
\label{eq:Rdyn-zm}
    \mathcal{R}_\mathrm{dyn}(z_m) &= p(M^i){\Delta_M}^i p({r_v}^j){\Delta_r}^j \nonumber \\ &\times \sum_l \mathcal{R}_\mathrm{GC}(\hat{z}(t_m + \tau_l))p(Z^k \mid \hat{z}(t_m + \tau_l)){\Delta_Z}^k,
\end{align}
where $t_m$ is the lookback time corresponding to redshift $z_m$, $\hat{z}$ is the cosmological function converting lookback time to redshift, $\tau_l$ is the time each BBH $l$ from the simulated clusters takes to merge, so that $z_\mathrm{form}  = \hat{z}(t_m + \tau_l)$, and we sum over all BBH mergers $l$ from the given cluster.
For consistency with GWTC-3~\citep{2021arXiv211103606T}, we assume the cosmological parameters from~\citet{2016A&A...594A..13P}, as implemented by \textsc{astropy}~\citep{2018AJ....156..123A}, for all cosmological calculations.
For a given GC population, the predicted BBH merger rate is summed over the prediction for each \cmc simulation (Eq.~\ref{eq:Rdyn-zm}) on the grid. We assume that the probability distributions $p(r_v)$ and $p(Z\mid z_\mathrm{form})$ only have support in the range covered by the grid. The case of the mass distribution $p(M)$ is discussed below.

\underline{Metallicities:} At each formation redshift, we assume a lognormal metallicity distribution, with the central metallicity given by Equation~6 in~\citet{2017ApJ...840...39M} and a metallicity spread of 0.5 dex. We assume solar metallicity $Z_\odot = 0.02$.

\underline{Masses:} We assume that the mass distribution follows a Schechter function between a minimum mass of $10^4\,\msun$ and a maximum mass of $10^8\,\msun$, with a power law slope $\beta_m$ and Schechter mass $M^\star$:
\begin{equation}
p(M) \propto \left(\frac{M}{M^\star}\right)^{\beta_m} \exp\left(-\frac{M}{M^\star}\right).
\end{equation} 
Unless we explicitly fit for $\beta_m$ and $M^\star$, our default values are $\beta_m = -2$ and $M^\star = 10^{6.3}\,\msun$ \citep{PortegiesZwartMcMillan2010,2020PhRvD.102l3016A}. The lower and upper bound of the mass distribution are degenerate with its shape and the GC formation rate. For simplicity, we fix these bounds and assume the mass distribution extends from $10^4$--$10^8\,\msun$, similar to the present-day masses of Milky Way GCs. However, our GC models from the \cmc catalog only cover the range $1.2$--$9.6\times10^5\,\msun$. We follow a similar procedure to \citet{2020ApJS..247...48K} to extrapolate to GC masses outside the simulated range by assuming that, at a fixed radius, the number of mergers scales with initial GC mass as $M^{1.6}$. This scaling relation comes from~\citet{2020MNRAS.492.2936A}, who found that it applies to a wide range of GC masses from $10^2$--$10^8\,\msun$. We verify that this relation holds over the smaller range of simulated masses from the \cmc catalog. While we extrapolate the total number of BBH mergers, we do not attempt to fit their delay times outside the simulated GC mass range. In other words, our predictions for the shape of the merger rate as a function of redshift is determined by GCs in the mass range $1.2$--$9.6\times10^5\,\msun$, with the normalization determined by our extrapolation to the mass range $10^4$--$10^8\,\msun$. We do not expect this approximation to noticeably impact our conclusions {within their current statistical uncertainties}. Across the \cmc GC models, the delay time distribution {among BBH mergers} does not vary strongly with GC mass, covering the full range from $\sim 0.1$ to $14$ Gyr with mean delays of 2--3 Gyr regardless of the GC mass ({see Fig.~\ref{fig:delay-time-v-cluster-mass}}). {If one alternatively considers all BBH systems, regardless of whether or not they merge within a Hubble time, the dependence of delay times on cluster mass becomes more significant, with smaller clusters producing higher fractions of BBH systems that fail to merge in a Hubble time (see, e.g., Fig. 1 of~\citealt{2016PhRvD..93h4029R}).}

\underline{Virial radii:} For the radius distribution, we take a Gaussian with mean $\mu_r$ and standard deviation $\sigma_r$, truncated between $0.5$ pc and $4$ pc. Unless we explicitly fit for the radius distribution, our default distribution is relatively flat in radius, with $\mu_r = 2$ pc, $\sigma_r = 2$ pc.

\underline{Formation redshift:} Within the mass and radius range considered, we assume that the cluster formation rate number density, $\mathcal{R}_\mathrm{GC}\left(z\right) \equiv \frac{dN_\mathrm{GC}}{dV_c dt} \left(z\right)$, follows a Madau-like function~\citep{2014ARA&A..52..415M}, which roughly follows $(1 + z)^{a_z}$ at low-$z$, peaks at $z_\mathrm{peak}$, and then follows $(1 + z)^{-b_z}$ at $z > z_\mathrm{peak}$:
\begin{equation}
\label{eq:cluster-formation-param}
    \mathcal{R}_\mathrm{GC}(z) = \mathcal{R}_0 \frac{(1 + z)^{a_z}}{1 + [(1 + z)/(1 + z_\mathrm{peak})]^{a_z + b_z}}
\end{equation}
We assume a maximum cluster formation redshift of $z_\mathrm{max} = 20$, above which we set $\mathcal{R}_\mathrm{GC}( z > z_\mathrm{max}) = 0$. 

We define the time integral of Eq.~\ref{eq:cluster-formation-param}, 
\begin{equation}
\label{eq:n0-def}
n_0 \equiv \int_0^{t_\mathrm{max}}\mathcal{R}_\mathrm{GC}(t)dt,
\end{equation}
where $\mathcal{R}_\mathrm{GC}(t) = \mathcal{R}_\mathrm{GC}(\hat{z}(t))$ and $t_\mathrm{max}$ is the lookback time corresponding to the maximum redshift $z_\mathrm{max} = 20$. 
If $n_0$ is known, then one can solve for $\mathcal{R}_0$ in Eq.~\ref{eq:cluster-formation-param} for any choice of $a_z$, $z_\mathrm{peak}$ and $b_z$. 
We can use the observed cluster number density to infer $n_0$. If all clusters that form over the history of the Universe survive to the present day, then the present-day cluster number density is simply $n_0$. However, many clusters do not survive to the present day, or else lose a substantial fraction of their stars through evaporation~\citep{2018PhRvL.121p1103F,2019ApJ...873..100C,2020PhRvD.102l3016A}. The observed number density of clusters today is therefore a lower limit on $n_0$, defined in Eq.~\ref{eq:n0-def}, and the true GC formation rate can be a factor of $\mathcal{O}(10)$ times higher~\citep{2020PhRvD.102l3016A}. We treat this cluster evaporation multiplicative factor as a free parameter in our model, defined as:
\begin{equation}
    f_\mathrm{ev} \equiv n_0 / n_\mathrm{surv},
\end{equation}
where $n_\mathrm{surv}$ is the observable present-day number density of survived clusters, $n_\mathrm{surv} = n_0/f_\mathrm{ev}$. 
We assume $n_\mathrm{surv} = 2.31 \times 10^9$ Gpc$^{-3}$ and a flat prior on $\log_{10} f_\mathrm{ev}$ between 0 and 2~\citep{2000ApJ...528L..17P,2020ApJS..247...48K,2020PhRvD.102l3016A}. 
This choice for $n_\mathrm{surv}$ applies to the Milky Way GC population. By restricting our assumed birth mass distribution to be between $10^4$ and $10^8\,\msun$, we cover a similar mass range to the present-day Milky Way GCs. For our default mass distribution, the average cluster birth mass is $4.0\times10^5\,\msun$, so that the GC mass density corresponding to our assumed number density is $\rho_\mathrm{surv} = 9.2\,\msun \times 10^{14}\,\msun$ Gpc$^{-3}$, which is well within the uncertainty interval found by~\citet{2020PhRvD.102l3016A}; see their Eq. 2.

Our treatment of cluster evaporation is a simplification. In reality, all clusters experience mass loss, with the smallest ones fully disrupting, causing both the GC number density and mass distribution to evolve with time. We do not model this full physical picture in our analysis. 
Instead, the free parameter $f_\mathrm{ev}$ captures the effects of cluster evaporation on the BBH merger rate alone, and is simply a way to parametrize the uncertain normalization of the GC formation rate of Eq.~\ref{eq:cluster-formation-param}.
The full set of hyper-parameters describing the GC population is $\{\beta_m, M^\star, \mu_r, \sigma_r, f_\mathrm{ev}, a_z, z_\mathrm{peak}, b_z\}$. 

\section{Merger rate evolution inferred from GWTC-3}
\label{sec:GW-fit}

\begin{figure}
    \centering
    \includegraphics[width = 0.5\textwidth]{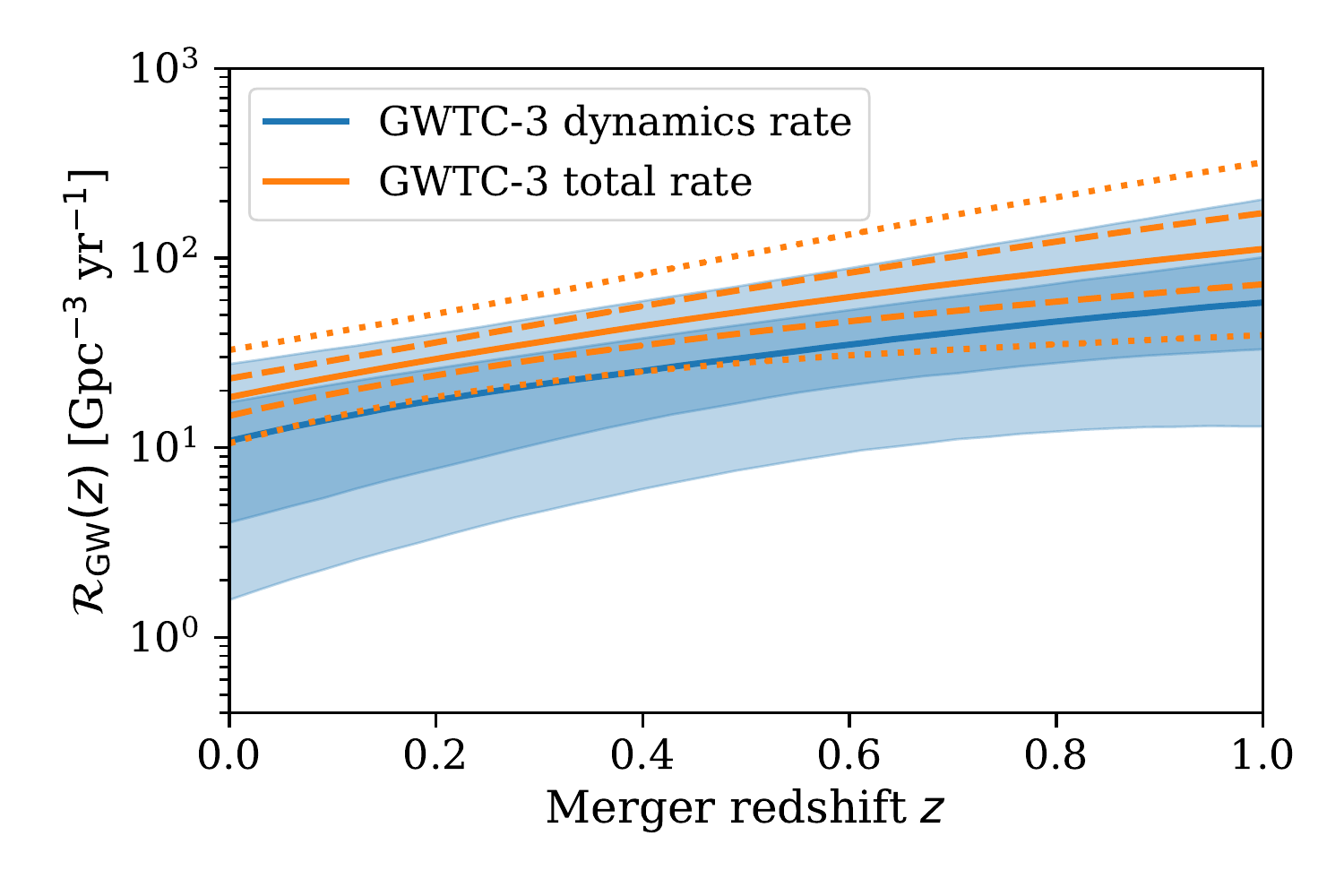}
    \caption{Merger rate as a function of redshift for BBH systems inferred from GWTC-3. Orange, unfilled bands show the total BBH merger rate, while the blue, filled bands show the dynamical contribution to the merger rate, inferred by measuring the fraction of BBH systems as a function of redshift that are consistent with an isotropic spin tilt distribution. Solid lines denote the median $\mathcal{R}(z)$ curve, while bands denote 50\% and 90\% credibility regions.}
    \label{fig:GWTC3_dyn_total_rates}
\end{figure}

\begin{figure}
    \centering
    \includegraphics[width=0.5\textwidth]{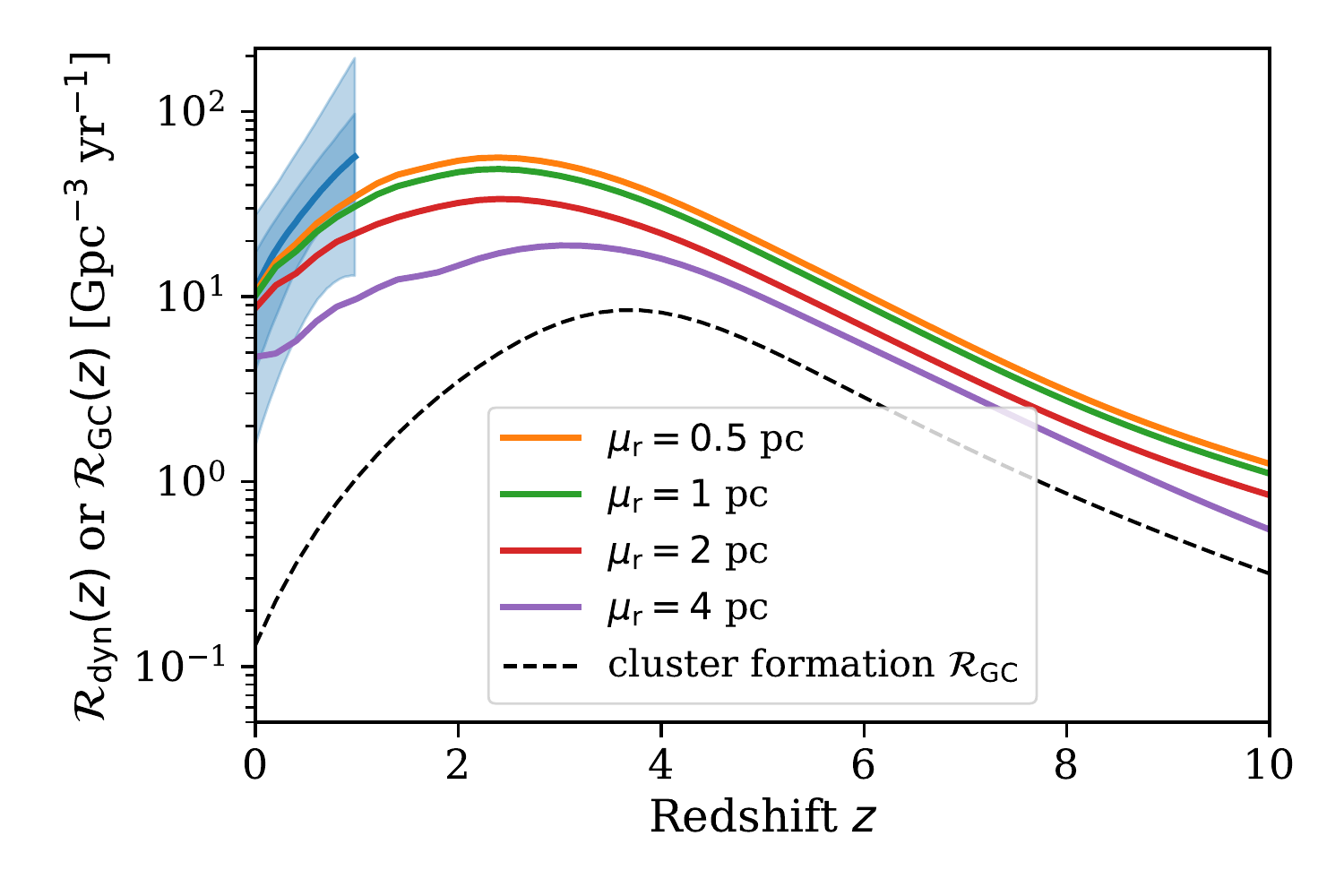}
    \caption{Inferred rate of dynamically-assembled BBH mergers (blue bands show 50\% and 90\% credible intervals) compared to the predicted rates from GCs with different radius distributions parameterized by $\mu_r$ (colored, solid lines). These BBH rate predictions assume that the formation rate density of GCs in the mass range $10^4$--$10^8\,M_\odot$ follows the dashed black line, with $f_\mathrm{ev} = 10$. For this plot, we fix the width of the radius distribution to $\sigma_r = 1$ pc, and GC birth mass distribution to the default Schechter function.}
    \label{fig:GWTC3-v-GCpred}
\end{figure}

Using GWTC-3 observations, we fit the BBH differential merger rate density as a function of masses, spins and merger redshift. We use a parameterized, phenomenological model to capture the main features of the BBH population; further details can be found in \S\ref{subsec:detailed-pop-models}.

Our goal is to measure the merger rate of BBH systems assembled in GCs as a function of redshift. We use the statistics of BBH spin tilts to distinguish between dynamically-assembled systems and those that evolved from other formation channels, such as isolated binary evolution~\citep{2016ApJ...832L...2R}. In particular, BBH systems originating from GCs are predicted to have randomly oriented (isotropic) spins, with no preference for alignment of the two component BH spins with each other or with the orbital angular momentum. The full BBH population, as inferred from GWTC-3, is inconsistent with an isotropic spin tilt distribution; there is a preference for systems with spins aligned to within 90 degrees of the orbital angular momentum axis. Therefore, the full BBH merger rate inferred from GWTC-3 cannot be attributed to GCs. Nevertheless, some fraction of systems are consistent with coming from an isotropic spin tilt distribution. The inferred merger rate of this isotropic subpopulation is an upper limit on the rate of BBH mergers originating in GCs. 

In order to measure the rate of BBH mergers with isotropic spin tilts, we fit the population distribution of effective inspiral spins $\chi_\mathrm{eff}$. The effective inspiral spin is defined as:
\begin{equation}
\chi_\mathrm{eff} = \frac{m_1 \chi_1 \cos t_1 + m_2 \chi_2 \cos t_2}{m_1 + m_2},
\end{equation}
where $m_1$ is the primary (bigger) mass, $\chi_1$ is the spin magnitude of the primary component, and $t_1$ is its spin tilt, and $m_2$, $\chi_2$ and $t_2$ describe the secondary (smaller) component. 
In contrast to the component spin magnitudes and tilts, $\chi_\mathrm{eff}$ is approximately conserved during the GW inspiral and is well-measured from the signal. A subpopulation with isotropic spin tilts (flat in $\cos t_1$ and $\cos t_2$) is symmetric about $\chi_\mathrm{eff} = 0$, with an equal fraction of systems having positive and negative $\chi_\mathrm{eff}$~\citep{2017Natur.548..426F}. In particular, the fraction of systems $f_\mathrm{neg}$ with $\chi_\mathrm{eff} \leq 0$, which is inferred to be smaller than the corresponding fraction $f_\mathrm{pos}$ with $\chi_\mathrm{eff} \geq 0$, gives an upper limit on the fraction of systems from a dynamically-assembled, isotropically spinning subpopulation, $f_\mathrm{dyn}$:
\begin{equation}
f_\mathrm{dyn} \leq 2f_\mathrm{neg} < 2f_\mathrm{pos}.
\end{equation}

We model the $\chi_\mathrm{eff}$ distribution at each redshift as a mixture between a Gaussian centered at zero, representing the ``dynamical" subpopulation, and a truncated Gaussian restricted to positive $\chi_\mathrm{eff}$. We refer to the fraction of systems in the zero-mean component as $f_\mathrm{dyn}(z)$, and allow both the overall merger rate and $f_\mathrm{dyn}(z)$ to evolve with redshift (similarly to~\citealt{2022A&A...665A..59B} and~\citealt{2022ApJ...932L..19B}; see \S\ref{subsec:detailed-pop-models}), thereby measuring the ``dynamics" rate as a function of redshift. 

Figure~\ref{fig:GWTC3_dyn_total_rates} shows our measurement of the total BBH merger rate (orange) and the merger rate of dynamically-assembled BBH (blue) as a function of redshift. The ``dynamics" rate should be thought of as an upper limit of the merger rate contributed by GCs, for two reasons. First, the mixing fraction $f_\mathrm{dyn}$ is technically an upper limit on the dynamically-assembled fraction. Other formation channels, including binary and triple star evolution in isolation or in young star clusters \citep[e.g.,][]{RodriguezAntonini2018,FragioneKocsis2020,2023arXiv230210851B}, may produce systems with $\chi_\mathrm{eff} < 0$, which would contribute to the zero-mean Gaussian component in our model. Second, in addition to GCs, other dense stellar environments, such as nuclear star clusters with or without an active galactic nucleus \citep[e.g.,][]{2009MNRAS.395.2127O,PetrovichAntonini2017,HoangNaoz2018,2021ApJ...923L..23W}, may contribute to the dynamically-assembled subpopulation. Nevertheless, in the following, we attribute the dynamics rate plotted in Fig.~\ref{fig:GWTC3_dyn_total_rates} to GCs, and use it to constrain properties of the GC population. 
When presenting our results, we highlight how our conclusions would change if the merger rate from GCs were significantly lower than our estimate. 

\section{Constraints on globular cluster properties}
\label{sec:GC-fit}

\begin{figure}
    \centering
    \includegraphics[width = 0.5\textwidth]{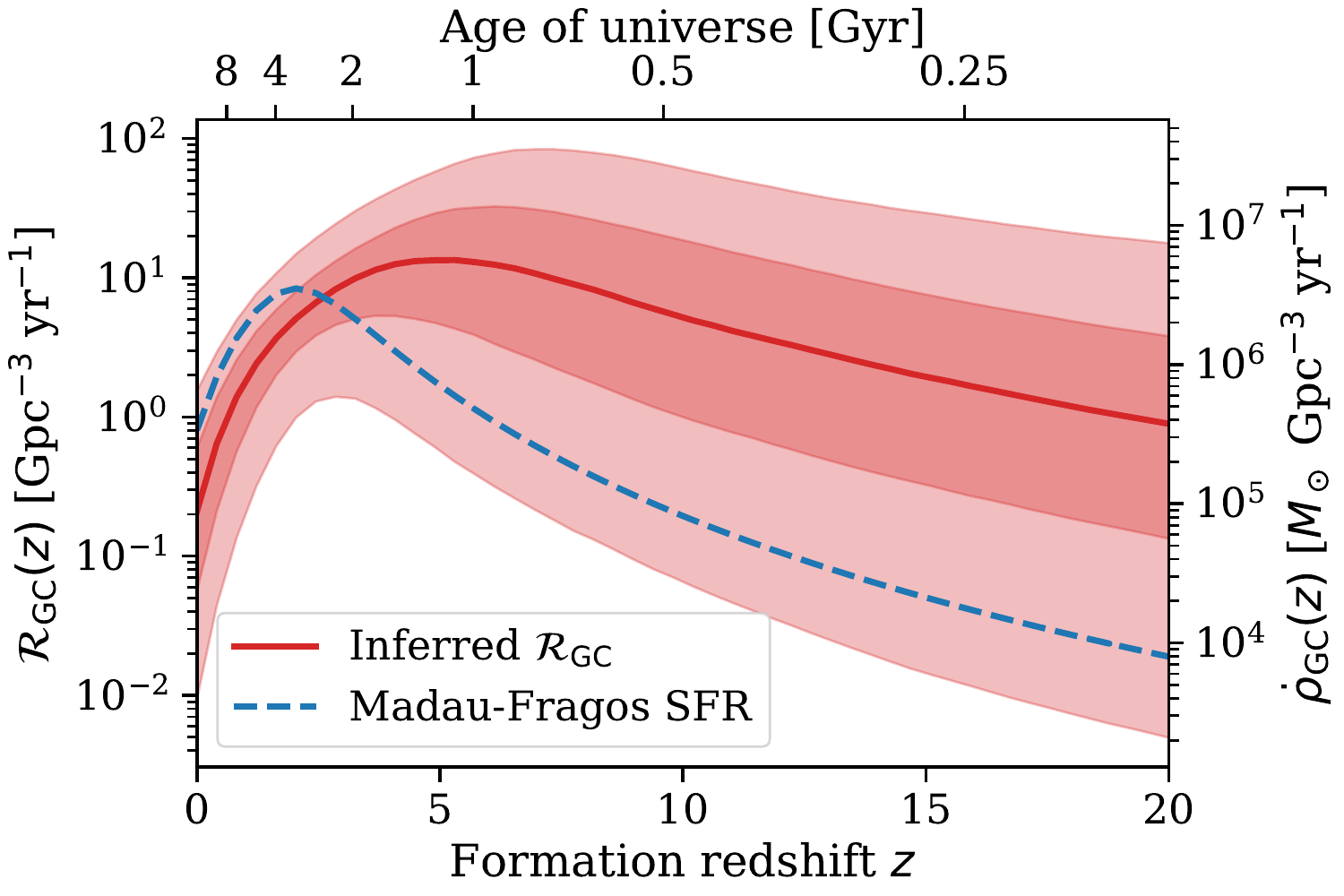}
    \caption{Cluster formation rate density $\mathcal{R}_\mathrm{GC}$ (left axis) and mass density $\dot{\rho}_\mathrm{GC}$ (right axis) as a function of redshift $z$ (bottom axis) and corresponding age of the Universe (top axis), as inferred from the GW ``dynamics" merger rate and the \cmc model suite (red). Solid line corresponds to the median $\mathcal{R}_\mathrm{GC}$ at each redshift, while shaded bands denote 50\% and 90\% credible intervals. {For reference, the blue, dashed line shows the SFR from~\citet{2017ApJ...840...39M} with an arbitrary normalization.}
    For this fit, we fix the GC mass function to the default Schechter function and marginalize over the radius distribution.}
    \label{fig:GCformation-inference}
\end{figure}

\begin{figure}
    \centering
    \includegraphics[width = 0.5\textwidth]{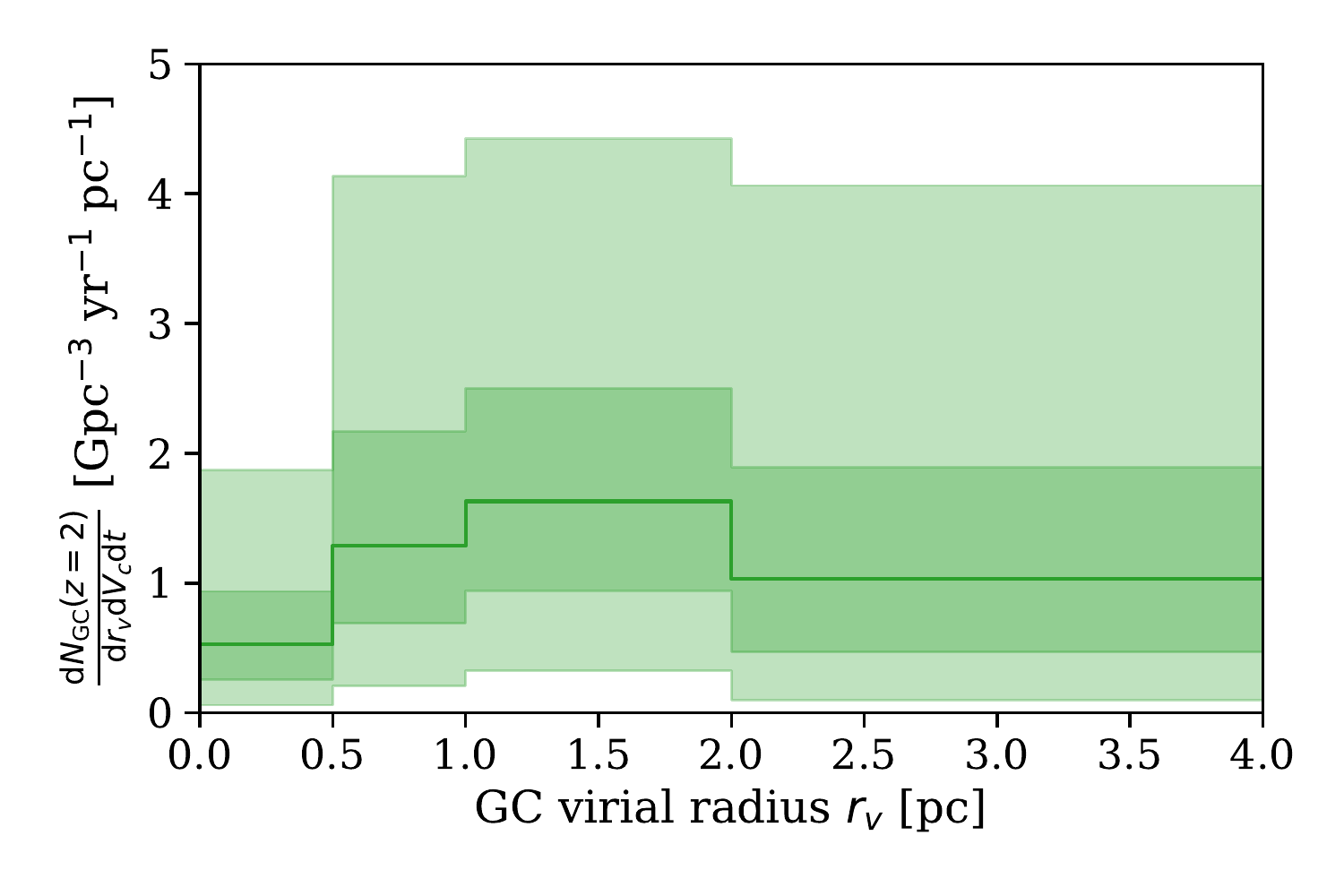}
    \caption{Distribution of GC virial radii $r_v$ as inferred from the GW ``dynamics" merger rate and the \cmc model suite, normalized so that integrating over $r_v$ gives the inferred GC formation rate density at $z  = 2$. We marginalize over the GC formation rate density $\mathcal{R}_\mathrm{GC}(z)$, and fix the GC birth mass distribution to the default Schecther function with $\beta_m = -2$. The discontinuous steps result from the distinct grid points at 0.5, 1, 2, and 4 pc from the \cmc model suite. The shape of the $r_v$ distribution is degenerate with the GC formation rate, and is not well constrained with current data.}
    \label{fig:GCradius-inference}
\end{figure}

\begin{figure}
    \centering
    \includegraphics[width = 0.5\textwidth]{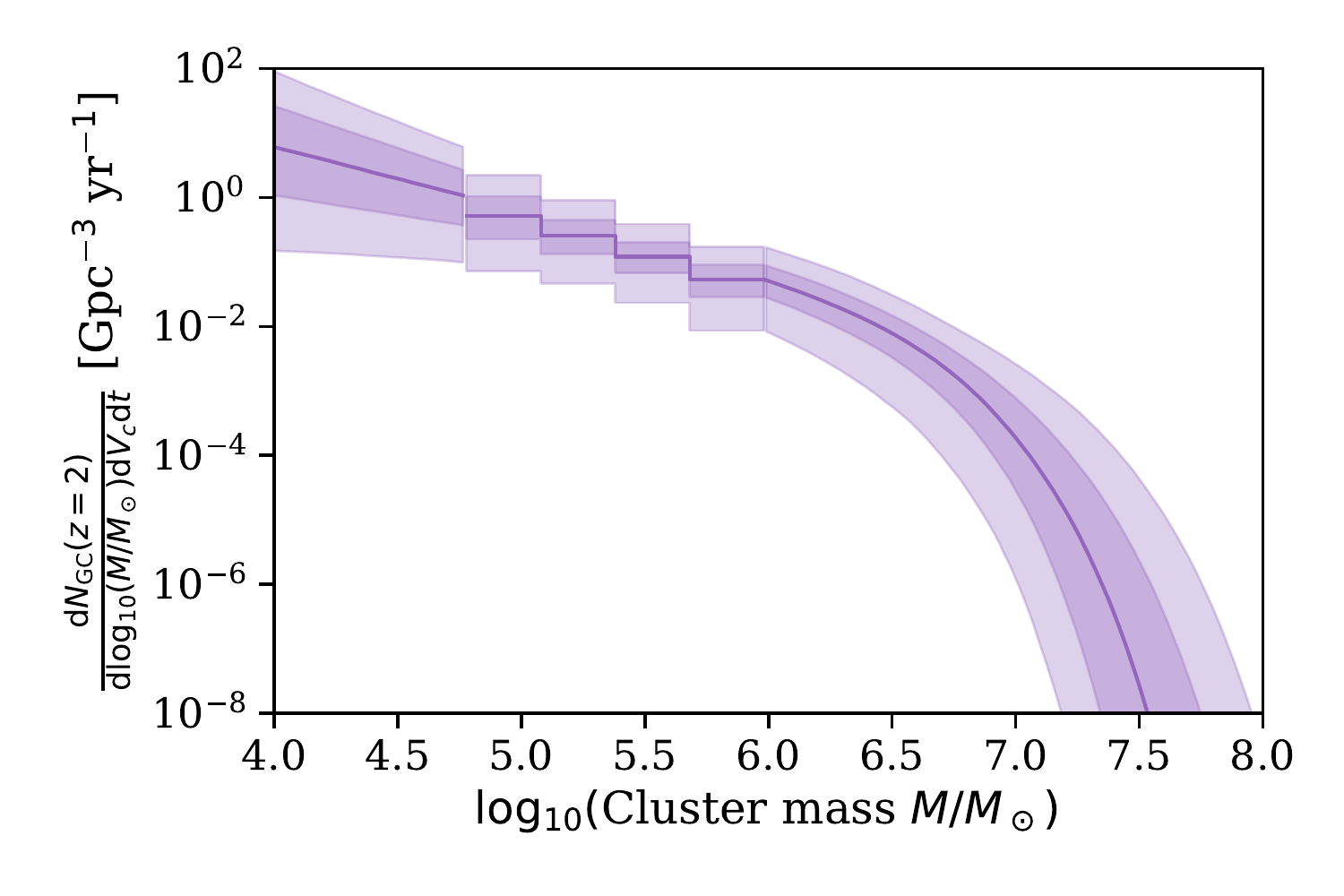}
    \caption{Distribution of GC $\log_{10}$ masses $\log_{10}M$ as inferred from the GW ``dynamics" merger rate and the \cmc model suite, normalized so that integrating over $\log_{10}M$ in the plotted mass range gives the inferred GC formation rate density at $z = 2$. We assume the mass distribution follows a Schechter function with variable power-law slope and Schechter mass, and marginalize over the GC formation rate density. We fix the GC radius distribution to the default, broad Gaussian centered at $\mu_r = 2$ pc with standard deviation $\sigma_r = 2$ pc. The discontinuous steps result from the distinct grid points, which we then extrapolate to cover the mass range $10^4$--$10^8\,\msun$.}
    \label{fig:GCmass-inference}
\end{figure}

Figure~\ref{fig:GWTC3-v-GCpred} shows the rate of dynamically-assembled BBH mergers measured from GWTC-3 overlaid with the predicted rates from a few GC models with varying virial radius distributions, fixing the formation rate density $\mathcal{R}_\mathrm{GC}$ (plotted as the dashed black line) and the initial mass distribution to the default Schechter function with slope $\beta_m = -2$ and Schechter mass $M^\star = 10^{6.3}\,M_\odot$.
The plotted formation rate $\mathcal{R}_\mathrm{GC}$ assumes a peak redshift of $z_\mathrm{peak} = 4$, low-$z$ slope of $a_z = 3$, high-$z$ slope of $b_z = 5$, and a cluster evaporation factor $f_\mathrm{ev} = 10$, leading to an effective present-day GC number density of $n_0/f_\mathrm{ev} = 2.31\times10^{9}$ Gpc$^{-3}$. 
As is well known, models that favor small radii and therefore denser clusters predict a higher BBH merger rate that increases slightly faster with redshift at $z \lesssim 2$ compared to those that favor more diffuse GCs \citep{2020ApJS..247...48K}. Similarly, a mass distribution that favors more massive clusters predicts a higher BBH merger rate, and varying the GC formation rate affects both the shape and amplitude of the BBH merger rate as a function of redshift. We therefore expect correlations in our inferred parameters describing the GC virial radius, mass, and formation redshift distribution. When we simultaneously fit all three distributions, these correlations make it difficult to interpret the results. Thus, in the following, we fix the GC mass distribution to the default Schechter function while jointly fitting for the radius distribution and the formation rate as a function of redshift. We then fix the radius distribution to the default Gaussian model and jointly fit for the initial mass function and formation redshifts. The redshift-dependent metallicity distribution is fixed to the~\citet{2017ApJ...840...39M} function in all calculations. 

In order to fit the distributions of GC virial radii, formation redshifts and masses to the GW data, we compare the measured BBH merger rate $\mathcal{R}_\mathrm{dyn}$ against GC predictions. We approximate the likelihood of the GW data given the distribution of GC properties; this approximate likelihood is based solely on the GW measurement of $\mathcal{R}_\mathrm{dyn}(z = 0)$ and $\mathcal{R}_\mathrm{dyn}(z = 1)$, similar to the method used in~\citet{2021ApJ...914L..30F}. In other words, our inference does not use any GC predictions for BBH masses, spin magnitudes, or eccentricities. This is an intentional, conservative choice to minimize modeling systematics that may affect GC predictions as well as the GW inference. More details are provided in \S\ref{subsec:approx-likelihood}. 
Using this approximate likelihood, we infer the hyper-parameters that describe the GC population, $\{\beta_m, M^\star, \mu_r, \sigma_r, f_\mathrm{ev}, a_z, z_\mathrm{peak}, b_z\}$. 

Figure~\ref{fig:GCformation-inference} shows our inferred GC formation rate as a function of redshift (red), assuming it follows the functional form of Eq.~\ref{eq:cluster-formation-param}, with flat priors on $a_z$, $z_\mathrm{peak}$, $b_z$, and a flat-in-log prior on $f_\mathrm{ev}$ ({see Fig.~\ref{fig:GC-formation-redshift-posterior-prior} for the posterior compared to the prior)}. The left axis shows $\mathcal{R}_\mathrm{GC}(z)$, or number of GCs in the mass range $10^4$--$10^8\,\msun$ formed per comoving volume and time, while the right axis shows the corresponding mass density $\dot{\rho}_\mathrm{GC}(z)$, or the total mass of GCs formed per comoving volume and time. For this analysis, we fix the mass distribution to the default Schechter function, and marginalize over the uncertain radius distribution. We find that the GC formation rate is best constrained around $\result{z = 2}$, with $\mathcal{R}_\mathrm{GC}(z = 2) = \result{5.0^{+9.4}_{-4.0}}$ Gpc$^{-3}$ yr$^{-1}$ in terms of number density, or $\dot{\rho}_\mathrm{GC}(z = 2) = \result{2.1^{+3.9}_{-1.7}}\times 10^6\,\msun$ Gpc$^{-3}$ yr$^{-1}$ in terms of mass density. If the true GC formation rate is lower, other environments must contribute to the ``dynamics rate" plotted in Fig.~\ref{fig:GWTC3_dyn_total_rates}. 

According to our fit, the GC formation rate likely rises more steeply than the SFR between redshift zero and $z \sim 3$. The \citet{2017ApJ...840...39M} SFR, {shown in the dashed blue curve of Fig.~\ref{fig:GCformation-inference}}, is $7.4$ times higher at $z = 3$ than $z = 0$, while our inferred GC formation rate is \result{$49.0^{+633.5}_{-46.6}$} times higher. This matches expectations that GC formation precedes the bulk of star formation, and GCs consist of older stellar populations. For example, the semi-analytic GC formation model of~\citet{2019MNRAS.482.4528E} predicts that GC formation peaked at $3 \lesssim z \lesssim 5$. While we do not measure $z_\mathrm{peak}$ well with available GW data (see Fig.~\ref{fig:corner-fixedmass} for the full posterior), our fit prefers $z_\mathrm{peak} \gtrsim 3$, and we rule out shallow low-$z$ slopes $a_z$ unless $z_\mathrm{peak}$ is very high. This inference is driven by the measured steep redshift evolution of the GW merger rate between $z = 0$ and $z = 1$, {which translates to a steep GC formation rate up to $z \sim 3$ because of the delay times that BBH mergers experience between formation and merger}.

Assuming a present-day GC number density of $2.31\times10^{9}$ Gpc$^{-3}$ for GCs within the initial mass range $10^4$--$10^8\,\msun$, we infer the evaporation factor $\result{f_\mathrm{ev} = 22.6^{+29.9}_{-16.2}}$. This is consistent with estimates based on observations of evolved GCs in the Milky Way. For example, \citet{2020PhRvD.102l3016A} infer that the initial GC mass density was 14.8--119.4 times higher than the present-day mass density, which, under their model, increases the BBH merger rate by a factor of 2.2--17.6 compared to the present-day density. Meanwhile, \citet{2018PhRvL.121p1103F} apply a factor of 2 to the predicted BBH merger rate to account for evaporation, and \citet{2019ApJ...873..100C} find a factor of 2.6. These estimates vary depending on the assumed initial GC mass function, including the lower and upper mass limits, and the details of GC evolution. Our inferred $f_\mathrm{ev}$ is comfortably higher than other estimates, given that it corresponds to the maximum possible contribution of GCs to the BBH merger rate. If other environments contribute to the ``dynamics" rate in Fig.~\ref{fig:GWTC3_dyn_total_rates}, then $f_\mathrm{ev}$ and the corresponding GC formation rate is lower than our reported values. {Conversely, if $f_\mathrm{ev}$ is known to be smaller, other scenarios likely contribute to the ``dynamics" rate. For example, if $f_\mathrm{ev} = 2$, GCs make up $8.9^{+22.5}_{-5.0}\%$ of the ``dynamics" rate (90\% credibility).}

Our fit to the GC virial radius distribution, assuming a Gaussian model with flat priors on the median $\mu_r$ and standard deviation $\sigma_r$, is shown in Fig.~\ref{fig:GCradius-inference}. This fit marginalizes over the GC formation history $\mathcal{R}_\mathrm{GC}(z)$, but fixes the default mass distribution. We plot the differential GC rate per virial radius at $z = 2$. The shape of the radius distribution $p(r_v)$ is assumed to be constant in redshift, but the normalization evolves with redshift. The discrete bins in Fig.~\ref{fig:GCradius-inference} correspond to the simulated grid of the \cmc catalog. We are not able to measure $\mu_r$ and $\sigma_r$ well, especially because of degeneracies with the formation history (see Fig.~\ref{fig:corner-fixedmass}). In particular, the inferred $\mu_r$ is degenerate with $f_\mathrm{ev}$. If the evaporation factor $f_\mathrm{ev}$ is small ($f_\mathrm{ev} \lesssim 10$), implying a lower GC formation rate, we infer smaller $\mu_r$ close to $0.5$ pc. As we saw in Fig.~\ref{fig:GWTC3-v-GCpred}, this is as expected because small $\mu_r$ implies a preference for more compact clusters in which the average merger rate per cluster is higher. 

Fixing the virial radius distribution (to the default, approximately-flat radius distribution) and marginalizing over the formation rate history, our inferred GC mass distribution is shown in Fig.~\ref{fig:GCmass-inference}. The bins between masses 1.2--$9.6\times10^5\,\msun$ designate the mass grid of the \cmc catalog, but we extrapolate the mass distribution between $10^4$ and $10^8\,M_\odot$ as described in \S\ref{sec:models}. In this plotted mass range, we measure a power law slope of \result{$\beta_m = -1.9^{+0.8}_{-0.8}$} and Schechter mass \result{$\log_{10}(M^\star/\msun) = 6.4^{+0.5}_{-0.4}$}. The Schechter mass is not tightly constrained within our prior (flat between $6 < \log_{10}(M^\star/\msun) < 7$), but the slope $\beta_m$ is well-measured, and agrees with our default value $\beta_m = -2$ inspired by measurements of young clusters \citep[e.g.,][]{ZhangFall1999,GielesLarsen2006,McCradyGraham2007}. 

Because the BBH merger rate depends sensitively on the cluster mass and the cluster formation rate, there are significant degeneracies between our inferred $\beta_m$ and $f_\mathrm{ev}$, and to a lesser extent, $M^\star$ (see Fig.~\ref{fig:corner-fixedradius}). Our measurement of $\beta_m$ is largely driven by our physically-motivated prior on $f_\mathrm{ev}$, which, combined with the measured GC number density today, limits the allowed GC formation rate. In particular, the GC number density today sets a lower limit on the GC formation rate. Shallow power law slopes $\beta_m \gtrsim -1$ result in a high average rate of BBH mergers per cluster, and given the present-day GC number density, are ruled out because they would over-predict the BBH merger rate. Steep power law slopes $\beta_m \lesssim -3$ would require $f_\mathrm{ev} > 100$ to produce a high enough merger rate that matches the inferred GW dynamics rate $\mathcal{R}_\mathrm{dyn}$. However, recall that the BBH merger rate from GCs may be lower than our measurement of $\mathcal{R}_\mathrm{dyn}$, so our reported upper limit on $\beta_m$ is more robust than the lower limit.

\section{Discussion}
\label{sec:discussion}

In the preceding sections, we measured the rate of BBH mergers that, based on their spin directions, are consistent with a GC origin  (\S\ref{sec:GW-fit}), and then compared this rate against predictions from the \cmc catalog to infer GC population properties (\S\ref{sec:GC-fit}). We fit for the GC virial radius distribution, formation rate as a function of redshift, and initial mass distribution. We did not consider other uncertain GC properties, and our fits to the GC population rely exclusively on their predicted BBH merger rates. 

Our work can be extended in several directions. When analyzing the GW data, we inferred the rate of BBH mergers with $\chi_\mathrm{eff} < 0$ to estimate the rate of mergers originating in GCs. This estimate is an improvement over simply comparing GC predictions to the total merger rate, which clearly overestimates the GC contribution. Our strategy also circumvents the need to directly model the contributions from additional formation channels and {to} introduce the corresponding modeling systematics~\citep{2021ApJ...910..152Z,2021PhRvD.103h3021W,2022MNRAS.511.5797M}. However, we may still be overestimating the contribution from GCs, because multiple other formation channels and environments can produce BBH mergers with $\chi_\mathrm{eff} < 0$. In the future, we can use more specific criteria to isolate the GC contribution to the merger rate, including, for example, the BBH population distribution of orbital eccentricity~\citep{2018PhRvD..97j3014S,2018PhRvD..98l3005R,2019ApJ...871...91Z,2021A&A...650A.189A}, which can be used to measure the fraction of BBHs assembled in GCs~\citep{2021ApJ...921L..43Z,2022ApJ...940..171R}, and the rate of hierarchical mergers~\citep{2018PhRvL.120o1101R,2021ApJ...915L..35K,2021NatAs...5..749G} as inferred from the distribution of spin magnitudes~\citep{2017ApJ...840L..24F,2020PhRvD.102d3002B,2022ApJ...935L..26F}. Given robust predictions for the BBH population distribution of eccentricity, spin magnitudes, and/or masses as a function of GC properties, we may incorporate these additional BBH observables into our inference, and improve our constraints on GC properties. 
For example, GC simulations provide a clear prediction for the BBH orbital eccentricity distribution~\citep{2018PhRvD..97j3014S}, although unfortunately eccentricity is not yet well-measured from GWs. With improvements in gravitational waveform modeling of eccentric and precessing BBH systems~\citep{2020PhRvD.101d4049L,2021PhRvD.103j4021N,2023MNRAS.519.5352R}, it may be possible to reconstruct the BBH eccentricity distribution, or at least estimate the fraction of eccentric systems and thereby measure the rate of BBH mergers that originate in GCs~\citep{2021ApJ...921L..43Z}.

On the GC modeling side, there are several uncertain GC properties that we did not vary in this work. These include the IMF~\citep{2021ApJ...907L..25W} and the initial binary fraction in massive stars \citep{GonzalezKremer2021}, which can significantly affect both the rates and typical masses of mergers. For example, GCs born with a top-heavy IMF produce a larger number of BBH mergers, while GCs born with a bottom-heavy IMF lead to fewer mergers. Furthermore, the slope of the IMF affects the BBH merger rate, and would be degenerate with the cluster size and mass if we included it as a free parameter. There may also be correlations between mass, radius, and formation redshift of GCs that we did not consider in this work~\citep[e.g.,][]{GielesBaumgardt2010}. In the future, especially once more GW observations are available out to higher redshifts, we can fit for these GC properties jointly with the radius distribution, mass distribution and formation histories considered here.

Finally, our calculations used limited external information about GCs. Although we sometimes fixed, e.g., the GC initial mass function to a default distribution, we used broad priors on all free hyper-parameters. In particular, we adopted a very broad prior on the GC formation history as a function of redshift, including its normalization through the uncertain evaporation factor. New galaxy simulations and electromagnetic observations {at high redshift, in particular with the recently-launched JWST \citep[e.g.,][]{2022ApJ...940L..53V,2022ApJ...937L..35M}}, are expanding our knowledge of GCs, which will inform the priors we adopt in future analyses. For example, if we have a motivated upper limit on the GC evaporation factor that is stricter than our generous upper limit of $f_\mathrm{ev} = 100$, we would deduce that the GC radius distribution likely favors smaller radii. If we had external knowledge that $f_\mathrm{ev} < 10$, we would conclude that other formation channels are probably contributing to the rate of BBH mergers with $\chi_\mathrm{eff} < 0$. By combining GW observations with other studies of GCs, we will arrive at a multi-messenger picture of GC formation and evolution.

\section{Conclusion}
\label{sec:conclusion}

In this work, we combined BBH merger rate predictions from the \cmc simulation catalog with GW observations in order measure the population properties of GCs: their virial radius distribution, formation history, and birth mass distribution. Our main results are as follows:
\begin{enumerate}
    \item By measuring the distribution of BBH spin orientations as a function of redshift, we infer that the rate of dynamically-assembled BBH mergers is $\result{10.9^{+16.8}_{-9.3}}$ Gpc$^{-3}$ yr$^{-1}$ at $z = 0$ and $\result{58.9^{+149.4}_{-46.0}}$ Gpc$^{-3}$ yr$^{-1}$ at $z = 1$, or $\result{61^{+29}_{-44}\%}$ of the total merger rate. 
    \item Marginalizing over the uncertain GC virial radius distribution and assuming GC masses follow a Schechter function with slope $\beta_m = -2$ and Schechter mass $M^\star = 10^{6.3}\,\msun$, we find that the GC formation rate is around $\result{5.2^{+9.4}_{-4.1}}$ Gpc$^{-3}$ yr$^{-1}$ at $z = 2$ (or $\dot{\rho}_\mathrm{GC}(z = 2) = \result{2.2^{+4.0}_{-1.7}}\times 10^6\,\msun$ Gpc$^{-3}$ yr$^{-1}$ in terms of mass density). We infer that the GC formation rate is \result{$50.5^{+630.1}_{-46.8}$} times higher at $z = 3$ than $z = 0$, evolving faster than the SFR, which is 7.4 times higher at $z = 3$ than today~\citep{2017ApJ...840...39M}.
    \item Fitting the GC mass distribution to a Schechter function and marginalizing over the uncertain formation history, we measure a power law slope of $\result{\beta_m = -1.9^{+0.8}_{-0.8}}$, consistent with the common assumption $\beta_m = -2$.  
\end{enumerate}
Our results represent the most comprehensive constraints on the GC population, including masses, virial radii and formation redshifts, from GW events so far. 
As new electromagnetic observations of clusters are providing an unprecedented view into their histories, current and upcoming GW observations provide a powerful, complementary probe of GC formation. 

\section*{Acknowledgements}

We are thankful to Fred Rasio, Kyle Kremer and Claire Ye for useful discussions, and to Sharan Banagiri and Szabolcs Marka for their helpful comments on the manuscript. G.F. acknowledges support by NASA Grant 80NSSC21K1722 at Northwestern University.
This material is based upon work supported by NSF LIGO Laboratory which is a major facility fully funded by the National Science Foundation.

\section*{Data Availability}

We have made use of gravitational-wave data from the~\citet{10.5281_zenodo.5117703,ligo_scientific_collaboration_and_virgo_2021_5546663,ligo_scientific_collaboration_and_virgo_2021_5636816} and globular cluster models from~\citet{2020ApJS..247...48K}. Our analysis code and data behind the figures is available at \url{https://github.com/mfishbach/dynamical-bbh-evolution}.


\bibliographystyle{mnras}
\bibliography{references}

\appendix

\onecolumn

\section{Fitting the BBH Population}
\label{sec:stat-appendix}

\subsection{Phenomenological Population Models}
\label{subsec:detailed-pop-models}

We parameterize the distribution of masses, effective inspiral spin, and merger redshift with the following model:
\begin{equation}
\label{eq:full-GW-model}
p(m_1, m_2, \chi_\mathrm{eff}, z \mid \mathbf{\Lambda}_m, \gamma, \mathbf{\Lambda}_\chi, \kappa) = p(m_1 \mid \mathbf{\Lambda}_m)p(m_2 \mid m_1, \gamma)p(\chi_\mathrm{eff} \mid z, \mathbf{\Lambda}_\chi)p(z \mid \kappa)
\end{equation}
We take the primary mass distribution to be a variation of the \textsc{Power Law + Peak} model~\citep{2018ApJ...856..173T}, but we model the low- and high-mass end of the mass spectrum with a high-pass filter $h(m_1)$ at the minimum mass and a low-pass filter $l(m_1)$ at the maximum mass~\citep{2020ApJ...899L...8F,2022ApJ...931..108F}:
\begin{equation}
h(m_1 \mid m_\mathrm{min}, \eta_h) = \left(\frac{m_1}{m_\mathrm{min}}\right)^{\eta_h} \left(1 + \left(\frac{m_1}{m_\mathrm{min}}\right)^{\eta_h}\right)^{-1}
\end{equation}
\begin{equation}
l(m_1 \mid m_\mathrm{max}, \eta_l) = \left(1 + \left(\frac{m_1}{m_\mathrm{max}} \right)^{\eta_l} \right)^{-1}
\end{equation}
The ``power law" component is given by:
\begin{equation}
p_\mathrm{PL}(m_1 \mid \alpha) = \frac{\alpha + 1}{100^{\alpha+1} - 2^{\alpha+1}}\left(\frac{m_1}{\msun}\right)^\alpha, 
\end{equation}
for $2\,\msun < m_1 < 100\,\msun$, and 0 elsewhere.
The ``peak" component is given by a Gaussian centered at $m_\mathrm{peak}$ with standard deviation $w_\mathrm{peak}$, which we denote as:
\begin{equation}
\mathcal{N}(m_1 \mid m_\mathrm{peak}, w_\mathrm{peak})
\end{equation}
Putting everything together, we model the primary mass distribution with eight parameters $\mathbf{\Lambda_m} = \{m_\mathrm{min}, m_\mathrm{max}, \eta_h, \eta_l, \alpha, f_\mathrm{peak}, m_\mathrm{peak}, w_\mathrm{peak}\}$:
\begin{equation}
p(m_1 \mid \mathbf{\Lambda_m}) = \left[(1 - f_\mathrm{peak})p_\mathrm{PL}(m_1 \mid \alpha) + f_\mathrm{peak}\mathcal{N}(m_1 \mid m_\mathrm{peak}, w_\mathrm{peak})\right]h(m_1 \mid m_\mathrm{min}, \eta_h)l(m_1 \mid m_\mathrm{max}, \eta_l).
\end{equation}
The secondary mass distribution, conditioned on the primary mass, is taken to be a power law with support between $2\,\msun$ and $m_1$ and slope $\gamma$:
\begin{equation}
p(m_2 \mid m_1, \gamma) = \frac{\gamma + 1}{m_1^{\gamma+1} - 2^{\gamma+1}}\left(\frac{m_2}{\msun}\right)^\gamma.
\end{equation}

For the spin distribution, we assume that at each redshift slice, $\chi_\mathrm{eff}$ follows a mixture model between two Gaussian components: a Gaussian centered at zero (``dynamics"), truncated to the physical range $-1 \leq \chi_\mathrm{eff} \leq 1$, and a Gaussian truncated to positive values $0 < \chi_\mathrm{eff} < 1$. 
The mixture fraction varies with redshift, so that the the redshift-dependent $\chi_\mathrm{eff}$ distribution is described by five parameters $\mathbf{\Lambda}_\chi = \{\sigma_\mathrm{dyn}, \mu_\mathrm{pos}, \sigma_\mathrm{pos}, f_\mathrm{dyn}^{z = 0}, f_\mathrm{dyn}^{z = 1} \}$:
\begin{equation}
\label{eq:pop-chieff-z}
    p(\chi_\mathrm{eff} \mid z, \mathbf{\Lambda}_\chi) = f_\mathrm{dyn}(z)\mathcal{N}^{T[-1,1]}(\chi_\mathrm{eff} \mid \mu = 0, \sigma = \sigma_\mathrm{dyn})
 + \left(1 - f_\mathrm{dyn}(z)\right)\mathcal{N}^{T[0,1]}(\chi_\mathrm{eff} \mid \mu = \mu_\mathrm{pos}, \sigma = \sigma_\mathrm{pos}),
\end{equation}
where 
\begin{equation}
\label{eq:fdyn-z}
    f_\mathrm{dyn}(z) = \left(1 + A\exp(k z)\right)^{-1},
\end{equation}
with $A = \frac{1}{f_\mathrm{dyn}^{z=0}} - 1$ and $k = \log \left(\frac{1}{f_\mathrm{dyn}^{z = 1}} -1 \right) - \log(A)$.

We model the evolution of the merger rate density with redshift as a power law in $(1 + z)$ with slope $\kappa$~\citep{2018ApJ...863L..41F}. Including the differential comoving volume $dV_c/dz$ and time dilation term $(1 + z)^{-1}$, the redshift distribution is therefore:
\begin{equation}
p(z \mid \kappa) = \frac{\frac{dV_c}{dz}(1 + z)^{\kappa - 1}}{\int_0^{z_\mathrm{max}} \frac{dV_c}{dz}(1 + z)^{\kappa - 1}dz},
\end{equation}
where $z_\mathrm{max}$ is the detection horizon (which we define based on the search sensitivity estimates;~\citealp{ligo_scientific_collaboration_and_virgo_2021_5636816}), and we compute $dV_c/dz$ for a Planck 2015 cosmology for consistency with GWTC-3~\citep{2016A&A...594A..13P, 2018AJ....156..123A}.

Equation~\ref{eq:full-GW-model} describes the normalized probability density. In addition to the hyper-parameters $\mathbf{\Lambda} \equiv \{\mathbf{\Lambda}_m, \gamma, \mathbf{\Lambda}_\chi, \kappa \}$, the normalization constant $N$ --- the total number of BBH mergers between $z = 0$ and $z_\mathrm{max}$, regardless of whether or not they are detected --- is also a free parameter in our model. $N$ is related to the merger rate density $\mathcal{R}_\mathrm{GW}(z)$ by:
\begin{equation}
    \mathcal{R}_\mathrm{GW}(z) = N \left(\frac{dV_c}{dz}\right)^{-1} (1 + z) p (z).
\end{equation}

\begin{table}
    \centering
    \begin{tabular}{ c  p{11cm} p{2mm} p{3cm} }
        \hline
        {\bf BBH hyper-parameter} & \textbf{Description} &  & \textbf{Prior} \\\hline\hline
        $m_\mathrm{min} / M_\odot$ & Low-mass end of the primary mass spectrum &  & U(3, 12)\\
        $m_\mathrm{max} / M_\odot$ &  High-mass end of the primary mass spectrum &  & U(50, 80)\\
        $\eta_h$ &  Smoothing parameter for the low-mass end of the primary mass spectrum & & U(1, 10) \\
        $\eta_l$ &  Smoothing parameter for the high-mass end of the primary mass spectrum &  & U(10, 20)\\
        $\alpha$ & Power-law slope of the primary mass distribution &  & U(-8, -0.5) \\
        $f_\mathrm{peak}$ & Fraction of primary masses in the Gaussian peak &  & U(0.001, 0.3) \\
        $m_\mathrm{peak}/\msun$ & Location of Gaussian peak &  & U(25, 50) \\
        $w_\mathrm{peak}/\msun$ & Width of Gaussian peak &  & U(2, 8) \\
        \hline
        $\gamma$ & Power-law slope of the secondary mass distribution &  & U($-2$, $8$) \\
        \hline
        $\sigma_\mathrm{dyn}$ & Width of zero-mean component in $\chi_\mathrm{eff}$ & & U(0.03, 0.5) \\
        $\mu_\mathrm{pos}$ & Location of positive component in $\chi_\mathrm{eff}$ & & U(0, 0.4) \\ 
        $\sigma_\mathrm{pos}$ & Width of positive component in $\chi_\mathrm{eff}$ & & U(0.03, 0.5) \\
        $f_\mathrm{dyn}^{z = 0}$ & Fraction of systems in zero-mean $\chi_\mathrm{eff}$ component at $z = 0$ & & U(0.01, 0.99) \\
        $f_\mathrm{dyn}^{z = 1}$ & Fraction of systems in zero-mean $\chi_\mathrm{eff}$ component at $z = 1$ & & U(0.01, 0.99) \\
        \hline
        $\kappa$ & Power-law slope in $(1 + z)$ of the merger rate evolution & & U(-8, 8) \\
        \hline
        $\log N$ & Normalization constant, total number of BBH mergers at $0 < z < z_\mathrm{max}$ & & U(7, 17) \\
        \hline
    \end{tabular}
    \caption{
    Summary of hyper-parameters $\{\mathbf{\Lambda}, N\}$ describing the phenomenological BBH population model in Eq.~\ref{eq:full-GW-model}.  The notation U$(a, b)$ denotes a uniform distribution between $a$ and $b$.
    }
  \label{tab:GWprior}
\end{table}

The likelihood for the GW data given the hyper-parameters $\{\mathbf{\Lambda}, N\}$ follows the standard hierarchical Poisson process likelihood, accounting for measurement uncertainty and observational selection effects~\citep{2019MNRAS.486.1086M}. To evaluate this likelihood, we use the same parameter estimation samples as~\citet{2021arXiv211103634T} for the 69 BBH events with a FAR $< 1$ yr$^{-1}$~\citep{2021SoftX..1300658A,10.5281_zenodo.5117703,ligo_scientific_collaboration_and_virgo_2021_5546663} as well as the search sensitivity estimates~\citep{ligo_scientific_collaboration_and_virgo_2021_5636816}. In the likelihood evaluation, we follow~\citet{2021arXiv210409508C} to convert the sampling priors on source parameters to the variables $m_1, m_2, z, \chi_\mathrm{eff}$.

The posterior on the hyper-parameters $\{\mathbf{\Lambda}, N\}$ are related to the likelihood by a prior.
We take broad, flat priors on all parameters in the set $\mathbf{\Lambda}$ and a flat-in-log prior on the normalization $N$. The prior ranges are listed in Table~\ref{tab:GWprior}. These priors play only a temporary role in the analysis. When we ultimately fit for the hyper-parameters describing the GC population, we ``un-do" the effect of these intermediate priors (see the following subsection \S\ref{subsec:approx-likelihood}). We sample from the posterior with \textsc{numpyro}~\citep{phan2019composable,bingham2019pyro}.

\subsection{Comparing to \cmc Models}
\label{subsec:approx-likelihood}

\begin{table}
    \centering
    \begin{tabular}{ c  p{8cm} p{2cm} p{2cm} p{2cm}}
        \hline
        {\bf GC hyper-parameter} & \textbf{Description} &  & \textbf{Prior} & \textbf{Default} \\\hline\hline
        $\beta_m$ & Power-law slope of Schechter function describing GC birth masses &  & U(-3, 3) & -2\\
        $\log_{10}(M^\star/\msun)$ & $\log_{10}$ Schechter mass of GC birth mass distribution &  & U(6, 7) & 6.3\\
        \hline
        $\mu_r/\mathrm{pc}$ & Mean of Gaussian describing virial radius distribution & & U(0.5, 4) & 2 \\
        $\sigma_r/\mathrm{pc}$ & Standard deviation of Gaussian describing virial radius distribution & & U(1, 3) & 2 \\
        \hline
        $\log_{10}(f_\mathrm{ev})$ & $\log_{10}$ of the ratio between the integrated GC formation rate $n_0$ and the GC density today $n_\mathrm{surv} = 2.31\times10^9$ Gpc$^{-3}$ & & U(0, 2) & 1 \\
        $a_z$ & GC formation rate roughly follows $(1+z)^{a_z}$ for $z < z_\mathrm{peak}$ & & U(1, 5) & 3 \\
        $z_\mathrm{peak}$ & Approximate peak redshift of GC formation rate & & U(0, 8) & 4 \\
        $b_z$ & GC formation rate roughly follows $(1+z)^{-b_z}$ for $z > z_\mathrm{peak}$ & & U(1, 5) & 5 \\
        \hline
    \end{tabular}
    \caption{
    Summary of hyper-parameters describing the GC population model detailed in \S\ref{sec:GC-param}. The notation U$(a, b)$ denotes a uniform distribution between $a$ and $b$. When we do not fit for a given hyper-parameter, we fix it to the default value listed in the last column.
    }
  \label{tab:GCprior}
\end{table}

\begin{figure}
\begin{minipage}[l]{0.47\textwidth}
    \includegraphics[width = \textwidth]{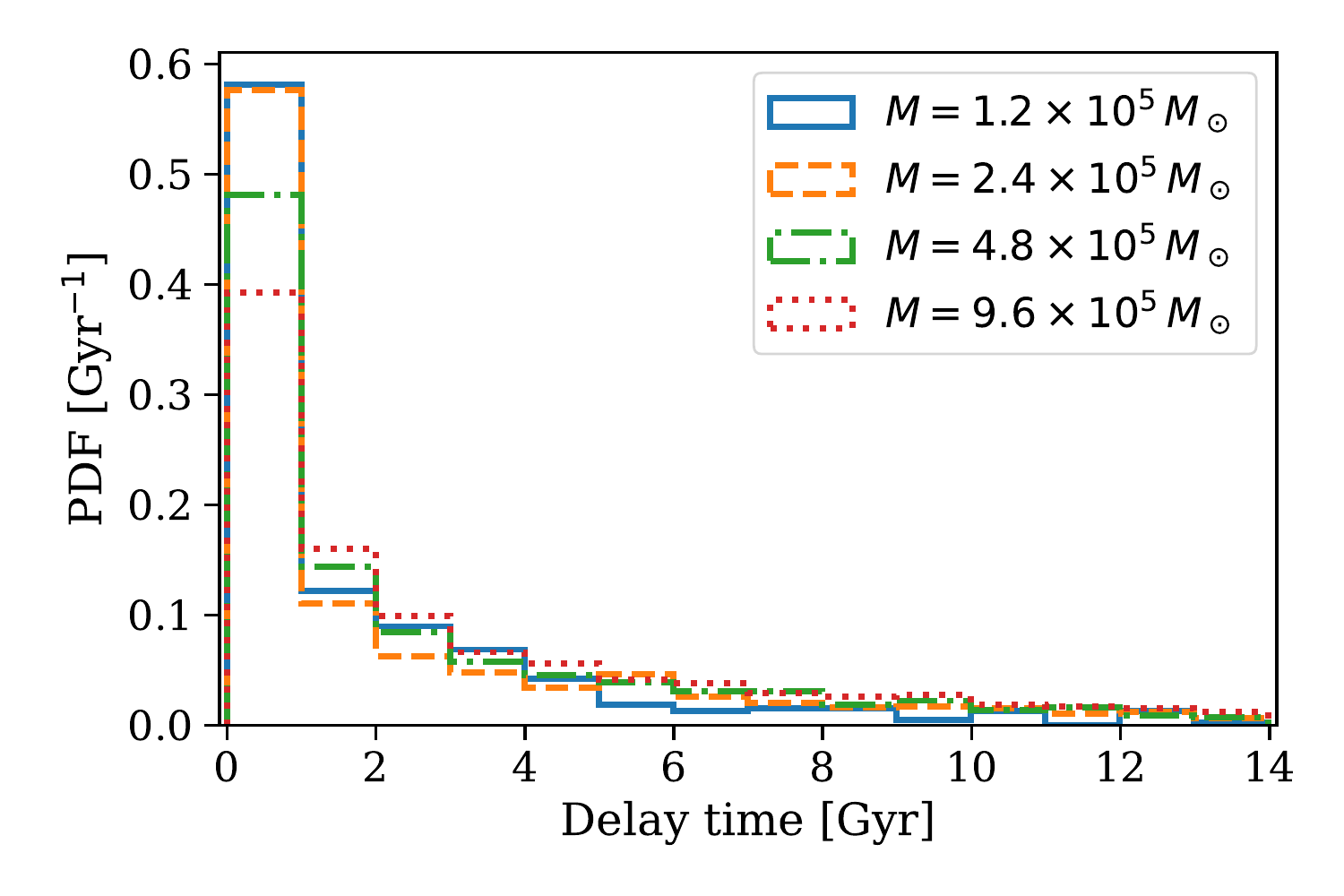}
    \caption{Probability density functions of the delay times between GC formation and BBH merger for GCs of different masses, as predicted by the \cmc catalog \citep{{2020ApJS..247...48K}}. For this plot, we weight GC virial radii and metallicities equally for each GC mass. We only consider BBH systems that merge within a Hubble time, so that averaging over GC radius, the delay time distributions are fairly similar across GC masses.}
    \label{fig:delay-time-v-cluster-mass}
\end{minipage}
\hspace{0.5cm}
\begin{minipage}[r]{0.48\textwidth} 
    \includegraphics[width = \textwidth]{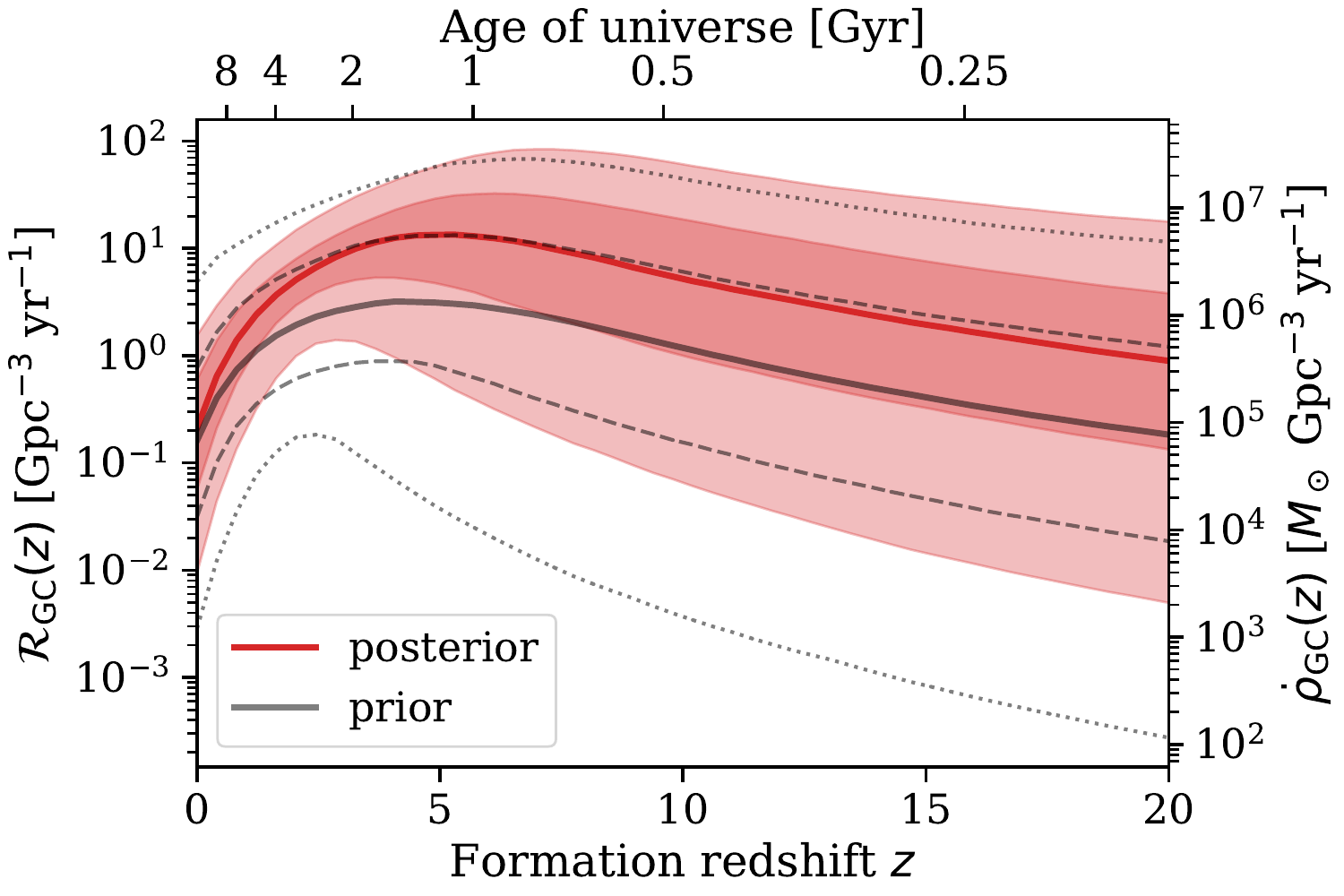}
    \caption{Inferred cluster formation rate as a function of redshift, as shown in Fig.~\ref{fig:GCformation-inference} (red), with the prior shown in gray. Solid lines show the median formation rate at each redshift. Dashed gray curves enclose 50\% of the prior probability and dotted gray curves enclose 90\% of the prior probability at each redshift. The data is informative about the shape of the GC formation rate at $z \lesssim 3$, and remains informative about the amplitude up to higher redshifts, although the shape of the GC formation rate at $z > 5$ is driven by the prior.}
    \label{fig:GC-formation-redshift-posterior-prior}
\end{minipage}
\end{figure}

In order to fit the GC mass, virial radius, and redshift distributions as parametrized in \S\ref{sec:GC-param}, we follow the method of~\citet{2021ApJ...914L..30F} to define an approximate likelihood for the GW data $d$ given the set of GC population hyper-parameters $\mathbf{\Theta} \equiv \{\beta_m, M^\star, \mu_r, \sigma_r, f_\mathrm{ev}, a_z, z_\mathrm{peak}, b_z\}$:
\begin{equation}
p(d \mid \mathbf{\Theta}) \approx p(d \mid \mathcal{R}_\mathrm{dyn}(z_m = 0 \mid \mathbf{\Theta}), \mathcal{R}_\mathrm{dyn}(z_m = 1 \mid \mathbf{\Theta})). 
\end{equation}
For each GC population described by $\mathbf{\Theta}$, we can calculate the predicted GC contribution to the BBH merger rate $\mathcal{R}_\mathrm{dyn}(z_m)$ at redshifts $z_m  = 0$ and $z_m = 1$ according to Eq.~\ref{eq:Rdyn-zm} {(see Fig.~\ref{fig:delay-time-v-cluster-mass} for the delay time distributions predicted by cluster models of different masses)}. 
Meanwhile, our fit to the BBH population described in \S\ref{sec:GC-fit} and \S\ref{subsec:detailed-pop-models} gives us draws from the BBH population hyper-posterior $p(\mathbf{\Lambda}, N \mid d)$, which we can transform into posterior draws on $\mathcal{R}_\mathrm{dyn}(z_m)$ for any merger redshift (see Fig.~\ref{fig:GWTC3_dyn_total_rates}), in particular redshifts 0 and 1. This gives us draws from the posterior density $p(\mathcal{R}_\mathrm{dyn}(0), \mathcal{R}_\mathrm{dyn}(1) \mid d)$.
Similarly, we draw from the BBH population hyper-prior specified in Table~\ref{tab:GWprior} and construct draws from the induced prior $p_0(\mathcal{R}_\mathrm{dyn}(0), \mathcal{R}_\mathrm{dyn}(1))$.
We apply a Gaussian kernel density estimate (KDE) to the posterior draws as well as the prior draws, which allows us to approximate the likelihood:
\begin{equation}
\label{eq:GC-approx-likelihood}
p(d \mid \mathbf{\Theta}) \approx p(d \mid p(\mathcal{R}_\mathrm{dyn}(0), \mathcal{R}_\mathrm{dyn}(1)) \propto \frac{ p(\mathcal{R}_\mathrm{dyn}(0), \mathcal{R}_\mathrm{dyn}(1) \mid d)}{p_0(\mathcal{R}_\mathrm{dyn}(0), \mathcal{R}_\mathrm{dyn}(1))}.
\end{equation}
Our posterior on the GC population hyper-parameters is given by the approximate likelihood in~Eq.~\ref{eq:GC-approx-likelihood} and the priors listed in Table~\ref{tab:GCprior}. {Figure~\ref{fig:GC-formation-redshift-posterior-prior} shows the influence of the prior on the inferred GC formation rate as a function of redshift.} We once again use \textsc{numpyro} to sample from the posterior~\citep{phan2019composable,bingham2019pyro}. 

\section{Full hyper-posteriors for cluster population fits}
\label{sec:corner-plots}
\begin{figure}
    \centering
    \includegraphics[width = \textwidth]{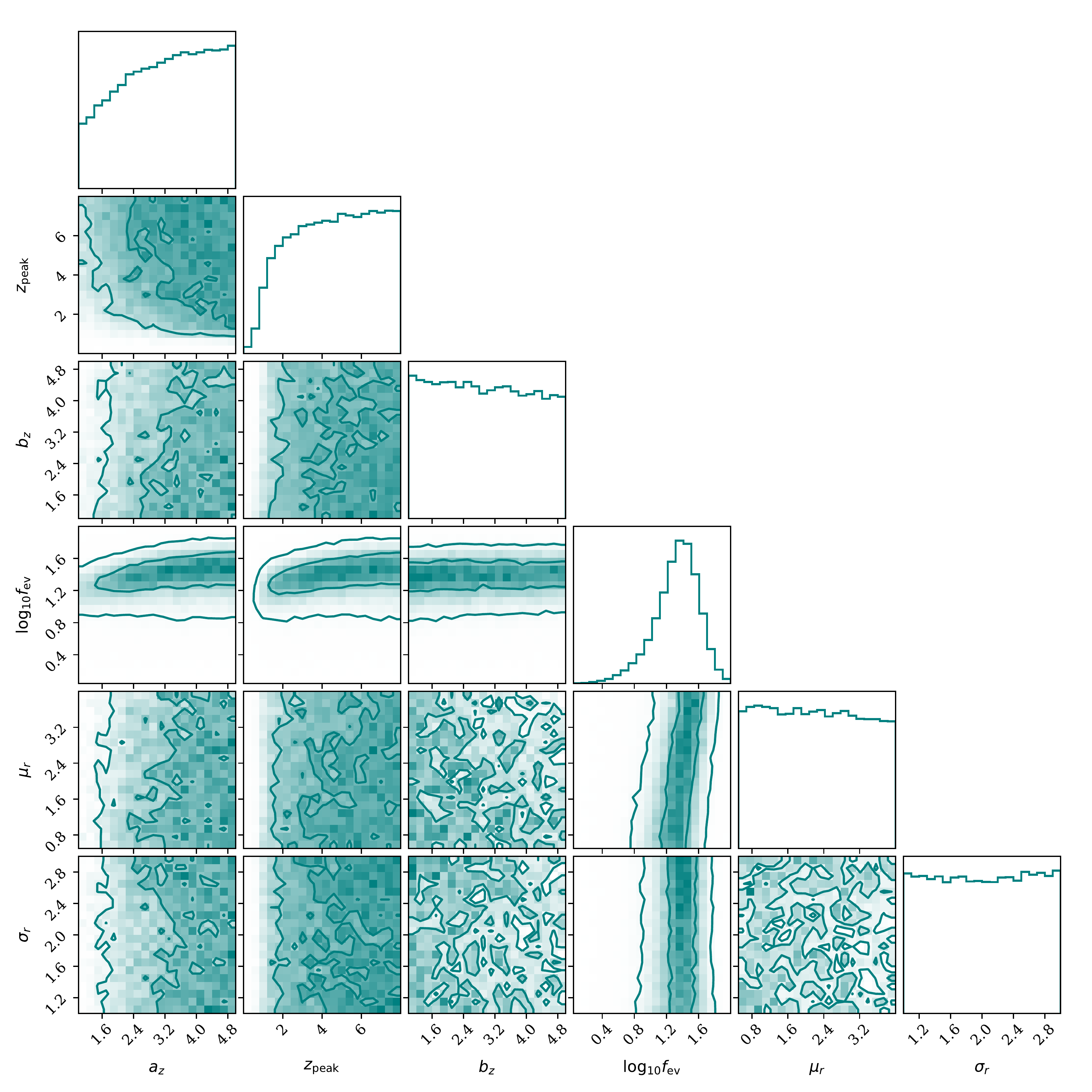}
    \caption{Hyper-posterior for parameters describing the GC virial radius and formation redshift distributions, fixing the mass distribution to the default values listed in Table~\ref{tab:GCprior}.}
    \label{fig:corner-fixedmass}
\end{figure}

\begin{figure}
    \centering
    \includegraphics[width = \textwidth]{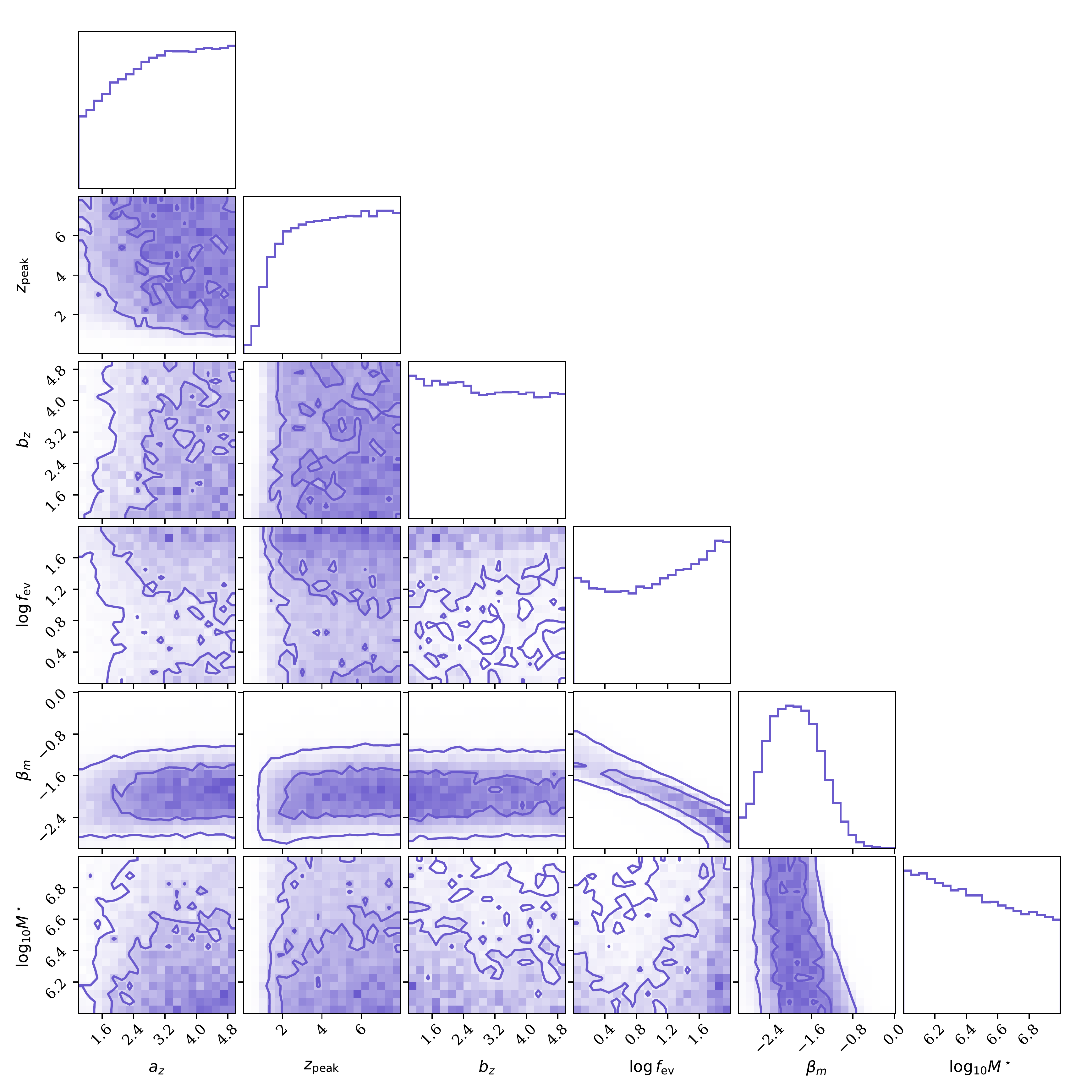}
    \caption{Hyper-posterior for parameters describing the GC birth mass and formation redshift distributions, fixing the virial radius distribution to the default values listed in Table~\ref{tab:GCprior}.}
    \label{fig:corner-fixedradius}
\end{figure}

We include corner plots summarizing the hyper-posteriors on the GC population parameters. Fig.~\ref{fig:corner-fixedmass} shows the inferred virial radius and formation redshifts hyper-parameters, fixing the mass distribution the default Schechter function, while Fig.~\ref{fig:corner-fixedradius} shows the inferred mass and formation redshift hyper-parameters, fixing the virial radius distribution to the default, relatively flat distribution (see Table~\ref{tab:GCprior} for priors and default values of all parameters).

\bsp	
\label{lastpage}
\end{document}